\documentclass[conference]{IEEEtran}
\makeatletter\if@twocolumn\PassOptionsToPackage{switch}{lineno}\else\fi\makeatother
\usepackage{bigdelim}

\usepackage{graphicx}
\usepackage[T1]{fontenc}
\usepackage[numbers,sort&compress]{natbib}
\usepackage{mathtools}
\usepackage{booktabs}
\usepackage{lineno,hyperref}
\usepackage{amsmath}
\usepackage{amsfonts}
\usepackage{amssymb}
\usepackage{amsmath}
\usepackage{multirow,morefloats,floatflt,cancel,tfrupee}
\usepackage{makecell}
\usepackage{tikz}
\usepackage{setspace}
\usepackage{color}
\usepackage{multirow}
\usepackage{adjustbox}
\usepackage{bbding}
\usepackage{physics}
\usepackage{colortbl}
\usepackage{xcolor}
\usepackage{pifont}
\usepackage{caption}
\usepackage[nointegrals]{wasysym}
\renewcommand\IEEEkeywordsname{Keywords}

\makeatletter
\def\ps@IEEEtitlepagestyle{%
  \def\@oddfoot{\mycopyrightnotice}%
  \def\@evenfoot{}%
}
\def\mycopyrightnotice{%
  {\footnotesize 979-8-3503-3015-1/23/\$31.00  \copyright 2023 IEEE \hfill}
  \gdef\mycopyrightnotice{}
}

\let\old@ps@IEEEtitlepagestyle\ps@IEEEtitlepagestyle
\def\confheader#1{%
    \def\ps@IEEEtitlepagestyle{%
        \old@ps@IEEEtitlepagestyle%
        \def\@oddhead{\strut\hfill#1\hfill\strut}%
        \def\@evenhead{\strut\hfill#1\hfill\strut}%
    }%
    \ps@headings%
}
\makeatother

\confheader{%
    \parbox{17cm}{\centering 13th International Conference on Computer and Knowledge Engineering (ICCKE 2023), November 1-2, 2023, Ferdowsi University of Mashhad}
}

\begin{document}

        \title{Capturing Local and Global Features in Medical Images by Using Ensemble CNN-Transformer}

\author{Javad Mirzapour Kaleybar$^{1}$
\quad Hooman Saadat$^{2}$
\quad Hooman Khaloo$^{3}$
\\
${^1}$ Department of Computer Engineering, University College of Nabi Akram, Tabriz, Iran  \\
${^2}$ Department of Electrical Engineering, Qazvin Branch, Iran\\
${^3}$ School of Technology Sharif University, Tehran, Iran\\
}

\maketitle

\begin{abstract}

This paper introduces a groundbreaking classification model called the Controllable Ensemble Transformer and CNN (CETC) for the analysis of medical images. The CETC model combines the powerful capabilities of convolutional neural networks (CNNs) and transformers to effectively capture both high and low frequency features present in medical images. The model architecture comprises three main components: a transformer classification block (TCB), a transposed-convolutional decoder block (TDB), and a convolutional encoder block (CEB).
The CEB is responsible for capturing multi-local features at different scales and draws upon components from VGGNet, ResNet, and MobileNet as backbones. By leveraging this combination, the CEB is able to effectively detect and encode local features. The TDB, on the other hand, consists of sub-decoders that decode and sum the captured features using ensemble coefficients. This enables the model to efficiently integrate the information from multiple scales. Finally, the TCB utilizes the SwT backbone and a specially designed prediction head to capture global features, ensuring a comprehensive understanding of the entire image.
The paper provides detailed information on the experimental setup and implementation, including the use of transfer learning, data preprocessing techniques, and training settings. The CETC model is trained and evaluated using two publicly available COVID-19 datasets. Moreover, the model is able to outperform existing state-of-the-art models across a variety of evaluation metrics, which is encouraging. Experimental results clearly demonstrate that the CETC model performs better than other deep learning models, emphasizing its potential for accurate and efficient image analysis in the field of medicine.
\end{abstract}


\section{Introduction}
\label{1}


Medical imaging relies on a diverse array of techniques such as optical coherence tomography, computed tomography, and magnetic resonance imaging to generate detailed visual representations of the body's internal structures \cite{saha2016active}. The interpretation of these medical images by healthcare professionals is of paramount importance for guiding future medical interventions. However, the process of training and analyzing the proficiency of human experts is time-consuming \cite{medvit}.
In the absence of skilled radiologists, advanced computer vision (CV) models have emerged as invaluable tools capable of swiftly providing highly accurate diagnostic assessments. Computer vision, as a prominent subset of deep learning field, encompasses a wide range of tasks including classification, segmentation, and denoising. Its overarching objective is to extract meaningful features from diverse visual inputs, encompassing photographs, videos, and various visual data sources.
These computer vision models offer the potential to revolutionize the field of medical imaging by expediting the diagnosis process, reducing human error, and improving patient outcomes. As technology continues to advance, the synergy between medical professionals and machine intelligence promises to enhance the efficiency and accuracy of healthcare diagnostics and treatment planning. The Transformer and CNN are two of the most commonly used classification models for images, both of which are used in image categorization.


CNNs, which have been a stalwart in the Computer Vision (CV) field for decades, excel at capturing hierarchical spatial information with additional convolutions \cite{robust, Embeded}. CNNs typically comprise convolution layers, dropout layers, fully connected layers, and other similar components.
In contrast, the transformer, a relatively newcomer to the CV landscape compared to CNNs, originally emerged within the realm of natural language processing and revolves around the concept of attention, also referred to as self-attention \cite{pyramid}. As elucidated by Saadati in \cite{Dilated}, attention involves the ability to connect various positions within a sequence to compute its representation. The pivotal distinction between CNN-based and transformer-based approaches lies in their focus: the former primarily consolidates local properties, whereas the latter takes a more global perspective \cite{Dilated}.


Our CETC architecture is structured around three fundamental components: a transformer classification block (TCB), a transposed-convolutional decoder block (TDB), and a convolutional encoder block (CEB). This amalgamation is designed to deliver a holistic feature extraction and classification process. Here's a detailed overview of how each element operates within CETC:
Convolutional Encoder Block (CEB): The CEB plays a pivotal role in the system by capturing features across a spectrum of local scales. Through the utilization of convolutional layers, it excels in discerning intricate patterns and details in the input data.
Transposed-Convolutional Decoder Block (TDB): Complementing the CEB, the TDB is responsible for decoding the captured features, returning them to their original scale. Subsequently, it integrates these decoded features using a set of three ensemble coefficients, thereby facilitating the effective amalgamation of multi-local information.
Transformer Classification Block (TCB): In the final phase of CETC, the aggregated features, now available at both multi-local and global scales, undergo further processing through transformer-based mechanisms. This phase is instrumental in extracting global features, which are essential for comprehensive understanding and classification.
The beauty of CETC lies in its seamless ability to capture features at both high and low frequency, thereby ensuring a comprehensive representation of the input data. Furthermore, our introduction of ensemble coefficients enables precise control over the contribution of local features across different scales, enhancing the model's adaptability and performance.
The robustness and effectiveness of CETC are evident in the evaluation results obtained from two prominent datasets: there are two datasets that are known to be available in COVID-19 radiography /cite[covid19radio, wang2020covid] and COVIDx CXR-3 \cite{covidxcxr}. It is consistently shown that CETC outperforms other models that are considered to be state-of-the-art (SOTA) across a wide range of evaluation metrics.
In summary, our primary contributions and innovations can be encapsulated as follows:

\begin{itemize}
\item The purpose of this study is to design a new CETC model that combines CNNs with transformers for the classification of medical images.
\item The developed CETC is capable of capturing both local and global features at the same time. At various scales, local features can be managed by adjusting the ensemble coefficients so that local contributions to the ensemble are optimized.
\item In our experiments on two open access datasets, we have found that our CETC excels SOTA methods in all evaluating metrics on both datasets.
\end{itemize}

The remaining sections of this paper are structured as follows: Section \hyperref[2]{2}, titled "Related Work," provides an overview of previously published studies on CNN and transformer models, specifically focusing on their applications in medical image classification. Section \hyperref[3]{3}, labeled "Materials and Methods," presents a summary of the datasets utilized in our research, as well as the implementation details, including architectural design, experimental settings, and evaluation metrics. In Section \hyperref[4]{4}, labeled "Results and Analysis," we present the outcomes of our study on two public COVID-19 datasets, along with a comprehensive analysis of the findings. Furthermore, we conducted additional ablation experiments. Finally, in Section \hyperref[5]{5}, entitled "Conclusions," we provide a summary of our work and discuss future research directions.

\section{The Proposed Approach}

\begin{figure*}
	\centering
	  \includegraphics[width=0.62\textwidth]{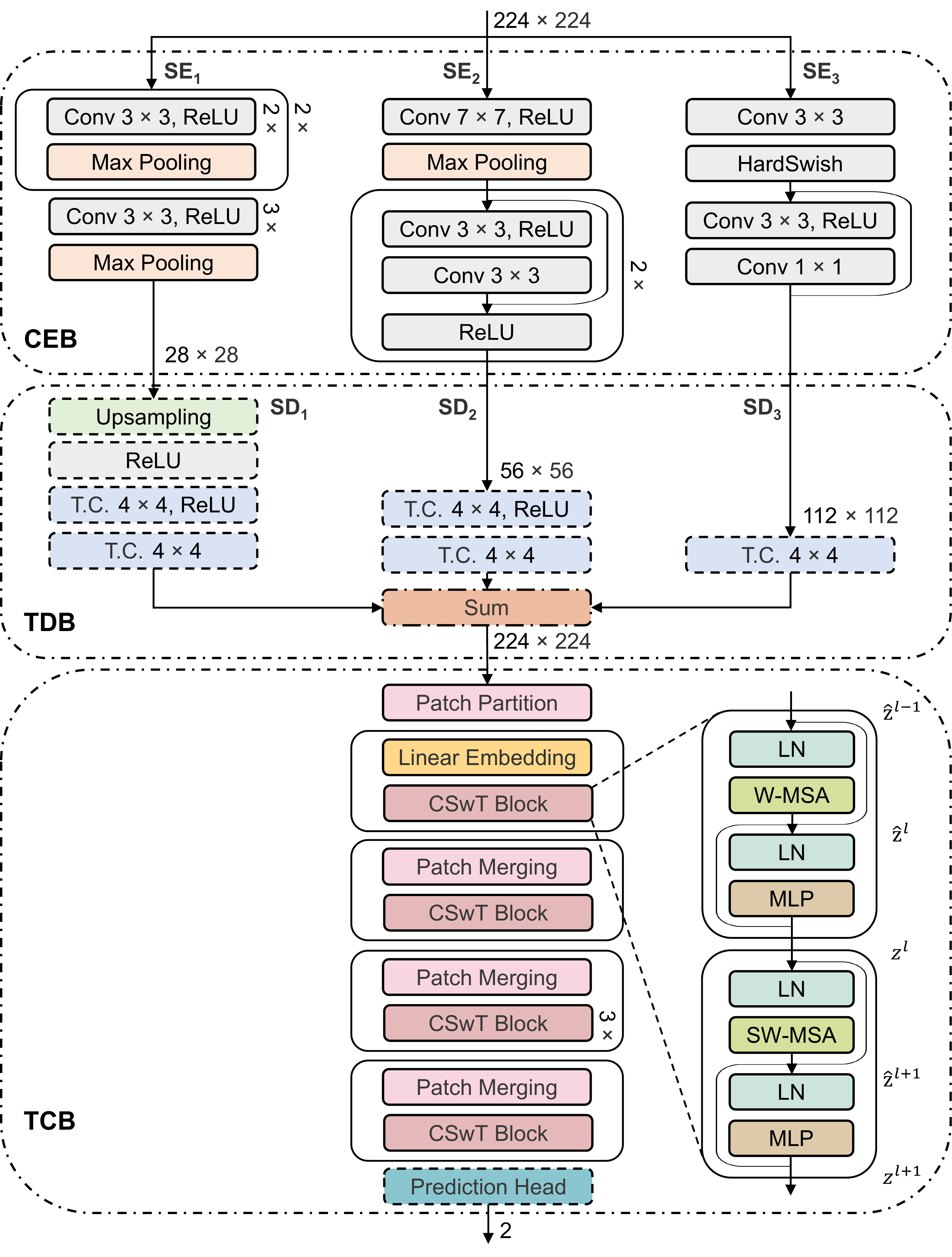}
	  \caption{The CETC architecture comprises several components. The CEB (Composite Encoder Block) consists of three sub-encoders named $\mathrm{SE}_1$, $\mathrm{SE}_2$, and $\mathrm{SE}_3$. On the other hand, the TDB (Triple Decoder Block) is composed of three sub-decoders denoted as $\mathrm{SD}_1$, $\mathrm{SD}_2$, and $\mathrm{SD}_3$. Within the TCB (Transformer Cross Attention Block), there are multiple CSwT (Cross-Shifted Windowing Transformer) blocks. SW-MSA (Shifted Windowed Multi-Head Self-Attention) and W-MSA (Windowed Multi-Head Self-Attention) modules represent multi-head attention modules that have regular and shifted windowing configurations, respectively, representing attention modules that focus on several heads simultaneously.}
	  \label{fig2}
\end{figure*}

The CETC model's detailed architecture is illustrated in Figure \ref{fig2}. It comprises three sub-encoders, denoted as $\mathrm{SE}_1$, $\mathrm{SE}_2$, and $\mathrm{SE}_3$. Each of these sub-encoders leverages components from well-established neural network architectures such as VGGNet \cite{simonyan2015very}, ResNet, and MobileNet as their core frameworks. To ensure uniformity with the $224 \times 224$ mount high level features, the output scales of $\mathrm{SE}_1$, $\mathrm{SE}_2$, and $\mathrm{SE}_3$ are set to $112 \times 112$, $28 \times 28$, and $56 \times 56$  respectively. This design choice allows for the capture of local features at different scales, specifically $112 \times 112$, $28 \times 28$, and $56 \times 56$ scales.
Moving on to the Transposed-Convolutional Decoder Block (TDB), it incorporates three sub-decoders denoted as $\mathrm{SD}_1$, $\mathrm{SD}_2$, and $\mathrm{SD}_3$, each specifically designed with an output dimension of 224 × 224 to match the input dimensions of $\mathrm{SE}_1$, $\mathrm{SE}_2$, and $\mathrm{SE}_3$. To consolidate the features produced by these sub-decoders, we employ an innovative feature map-level summation operation, utilizing the ensemble coefficients we have introduced.
Lastly, the TCB employs a Swin Transformer (SwT) as its core architecture, complemented by a specially tailored prediction head. This block takes the summed feature map as input and is proficient in capturing the global features at the 224 × 224 scale. Through the synergistic combination of the CEB, TDB, and TCB, the CETC model excels in effectively capturing features spanning from 28 × 28 to 224 × 224 scales, encompassing both local and global aspects.

In the CEB, $\mathrm{SE}_1$ originates from the upper 17 sub-layers of VGGNet, whereas $\mathrm{SE}_2$ and $\mathrm{SE}_3$ are derived from the upper 5 and 2 sub-layers of MobileNet and ResNet, respectively.


The TDB comprises distinct elements for each sub-decoder. $\mathrm{SD}_1$ encompasses an upsampled convolutional layer, followed by a ReLU filter function. It also encompasses a $5 \times 5$ transposed convolutional layer with an accompanying ReLU activation function, and a subsequent $5 \times 5$ transposed convolutional layer. $\mathrm{SD}_2$ comprises of a $5 \times 5$ transposed convolutional layer, succeeded by a ReLU activation function, and another $5 \times 5$ transposed convolutional layer. On the other hand, $\mathrm{SD}_3$ consists of a solitary 4 × 4 transposed convolutional layer. The process of feature map-level summation is carried out with the involvement of ensemble coefficients denoted as $\alpha$, $\beta$, and $\gamma$, and the resultant output $y$ is computed as follows:

\begin{equation}
y= \beta {FSD_2}+ \alpha {FSD_1}+\gamma {FSD_3},
\label{eq1}
\end{equation}

Here, $\alpha$, $\beta$, and $\gamma$ are coefficients ranging from 0 to 1, and their sum equals 1. ${FSD_2}$, ${FSD_1}$, and ${FSD_3}$ represent the output feature maps of $\mathrm{SD}_2$, $\mathrm{SD}_1$, and $\mathrm{SD}_3$, respectively.


In the TCB, a total of four levels are present, each featuring a different count of CSwT blocks. Each CSwT block is comprised of two components: one featuring a multi-head attention module employing standard windowing configurations (W-MSA), and the other employing shifted windowing configurations (SW-MSA). The MLP module consists of two layers, separated by a GELU activation function. The prediction head encompasses a linear layer with two output forks. The operation of the CSwT block can be summarized as follows:

\begin{equation}
\hat{\mathbf{z}}^l=\mathrm{W}-\mathrm{MSA}\left(\mathrm{LN}\left(\mathbf{z}^{l-1}\right)\right)+\mathbf{z}^{l-1},
\label{eq2}
\end{equation}

\begin{equation}
\mathbf{z}^l=\operatorname{MLP}\left(\operatorname{LN}\left(\hat{\mathbf{z}}^l\right)\right)+\hat{\mathbf{z}}^l,
\label{eq3}
\end{equation}

\begin{equation}
\hat{\mathbf{z}}^{l+1}=\operatorname{SW-MSA}\left(\mathrm{LN}\left(\mathbf{z}^l\right)\right)+\mathbf{z}^l,
\label{eq4}
\end{equation}

\begin{equation}
\mathbf{z}^{l+1}=\operatorname{MLP}\left(\operatorname{LN}\left(\hat{\mathbf{z}}^{l+1}\right)\right)+\hat{\mathbf{z}}^{l+1},
\label{eq5}
\end{equation}

Within this context, $\hat{\mathbf{z}}^l$ and $\mathbf{z}^l$ denote the output characteristics of the MLP and the W-MSA components within block $l$, respectively. It's worth noting that attention computations involve the inclusion of relative position bias.

\section{Experiments}

In this section, we initiate the discussion by examining the datasets utilized to train our enhanced Transformer model, and provide a summary of the experimental configurations. Following that, we conduct a comprehensive analysis of the experimental outcomes, here, by comparing the CETC model with the most recent advances in the task of the classification of medical images, we are able to assess the effectiveness of the model.

\subsection{Implementation details}

CETC is built using Pytorch deep learning and Pycharm IDE framework, which allows us to create our CETC in a very simple way. The input images undergo resizing and cropping to achieve a resolution of $224 × 224$, followed by normalization. To enhance the model's generalization ability, the training dataset's images are horizontally flipped. The loss function is determined by cross-entropy loss, and Adam is used as the optimization algorithm. For the CEB and TCB network components (excluding the prediction head), Transfer learning is used. Specifically, before being trained on the COVID-19 image datasets, these components are first pre-trained on the ImageNet dataset. We depict network components with transfer learning using solid lines in Figure \ref{fig2} and network components without transfer learning using dotted lines.

With a starting learning rate of $0.003$, we run training on all models for $20$ epochs. The reduction factor for the ReduceLROnPlateau function is $0.5$ and the patience factor for the ReduceLROnPlateau function is $5$ in order to progressively lower the learning rate as the training progresses. The batch size is set to $64$ to ensure that memory utilization and training time are balanced. We restrict the values of to keep the computational efficiency within a reasonable range $\beta$, $\alpha$, and $\gamma$ to be chosen from the set {0.1, 0.2, $0.\dot{3}$, 0.6, 0.8}, where $0.\dot{3}$ denotes $\frac{1}{3}$. Additionally, there must be at least 2 denominators that share the same merit. This restriction yields 7 groups of $\beta$, $\alpha$, and $\gamma$ combinations, as depicted in Table~\ref{tab2} or Table~\ref{tab3}.

\begin{figure*}
	\centering
	  \includegraphics[width=0.82\textwidth]{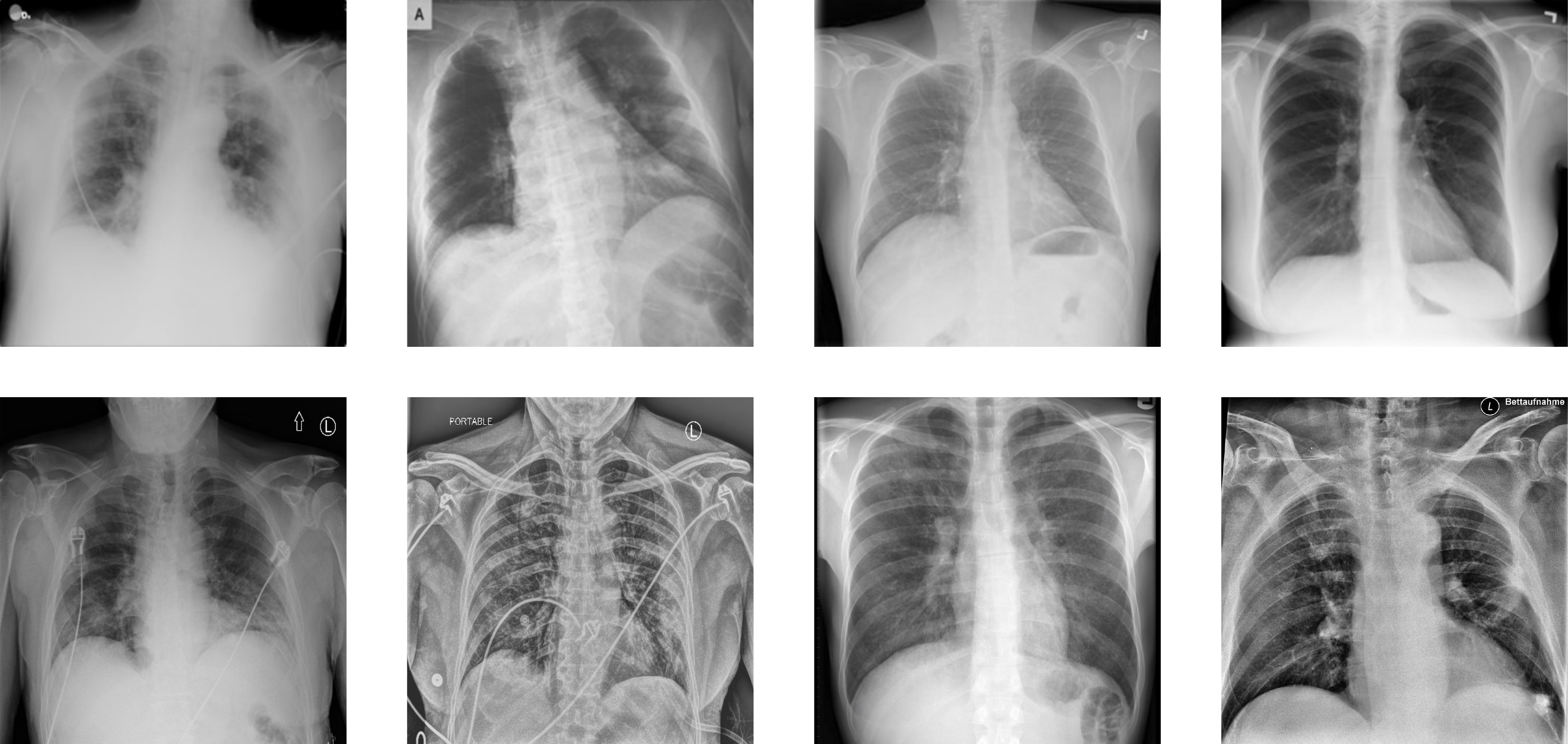}
	  \caption{As can be seen in each row, the first two images represent cases that test positive for COVID, while the last two images represent cases that test negative for COVID. The COVID-19 radiography dataset is exhibited on top of the photograph, whereas the COVIDx CXR-3 radiography dataset is exhibited on the bottom of the photograph.}
	  \label{fig1}
\end{figure*}

\subsection{Datasets}

We evaluated CETC's performance using two COVID-19 datasets that are available to the general public: the COVIDx CXR-3 dataset and the COVID-19 radiography dataset. Researchers from Qatar University, the University of Dhaka, and their partners created 3616 COVID-positive pictures and 10192 COVID-negative pictures that make up the COVID-19 radiography collection. The COVIDx CXR-3 dataset, on the other hand, consists of chest x-ray pictures gathered from 16,648 patients in various locales. This dataset consists of two sections that together include 30386 photos. There are 29986 photos in the first phase, which is utilized for training and validation, of which 15994 have been categorized as positive and 13992 as negative. 400 photos make up the second, testing portion, of which 200 are marked as positive and 200 as negative. Here are some sample images from both datasets for the purpose of reference in Figure~\ref{fig1}.

Each dataset was divided into several subgroups at random. We kept an 8:1:1 ratio while dividing the COVID-19 radiography dataset into training, validation, and test subsets. It was decided to split the COVIDx CXR-3 dataset into validation and training subsets using an 8:2 ratio because the testing images were provided individually, and it was therefore easiest for us to separate the dataset.

\subsection{Evaluation Measures}

For the goal of evaluating the efficiency of our model, we employ six evaluation metrics: positive predictive value (PPV), accuracy (ACC), negative predictive value (NPV), F-1 score (FOS), sensitivity (SEN), and specificity (SPE). These metrics can be computed using the following formulas, based on false negative (FN), the values of true positive (TP), false positive (FP), and true negative (TN),:

\begin{equation}
S E N=\frac{T P}{T P+F N},
\label{eq9}
\end{equation}

\begin{equation}
N P V=\frac{T N}{T N+F N},
\label{eq7}
\end{equation}

\begin{equation}
A C C=\frac{T P+T N}{T P+T N+F P+F N},
\label{eq6}
\end{equation}

\begin{equation}
P P V=\frac{T P}{T P+F P},
\label{eq8}
\end{equation}

\begin{equation}
S P E=\frac{T N}{T N+F P},
\label{eq10}
\end{equation}

\begin{equation}
F O S=\frac{2 T P}{2 T P+F N+F P},
\label{eq11}
\end{equation}

In the context of binary classification, the terms TP and TN refer to cases in which the model predicted the positive category correctly whereas the negative category correctly predicted by the model, respectively. On the other hand, FP stands for situations where a model incorrectly predicts the positive category, whereas FN stands for situations where the model incorrectly predicts the negative category. It is the TN, TP, FN, and  FP values that represent the four sub-blocks of the confusion matrix, which are indicated by their respective letters.

\subsection{Ensemble}

\begin{table}[ht]
\centering
\caption{A comparison of the CETC model under 7 diverse groups of $\beta$, $\alpha$, and $\gamma$ for the radiography dataset of COVID-19 with the performance of the CETC model. $0.\dot{3}$ represents $\frac{1}{3}$. It is essential to note that bolded outcomes are the best results.}
\begin{adjustbox}{width=.95\linewidth,center}
\begin{tabular}[t]{ccccccccc}
\toprule
    $\alpha$    & $\beta$     & $\gamma$    & ACC             & NPV             & PPV             & SEN             & SPE             & FOS    \\
\midrule
    0.8         & 0.1         & 0.1         & 98.0\%          & 98.8\%          & 95.9\%          & 96.7\%          & 98.5\%          & 96.3\% \\
    0.6         & 0.2         & 0.2         & 97.6\%          & \textbf{99.2\%} & 93.4\%          & \textbf{97.8\%} & 97.5\%          & 95.5\% \\
    0.1         & 0.8         & 0.1         & 97.8\%          & 98.6\%          & 95.3\%          & 96.1\%          & 98.3\%          & 95.7\% \\
    0.2         & 0.6         & 0.2         & 98.0\%          & 98.2\%          & \textbf{97.5\%} & 95.0\%          & \textbf{99.1\%} & 96.2\% \\
    0.1         & 0.1         & 0.8         & 97.5\%          & 98.6\%          & 94.6\%          & 96.1\%          & 98.0\%          & 95.3\% \\
    0.2         & 0.2         & 0.6         & 98.0\%          & 98.8\%          & 95.6\%          & 96.7\%          & 98.4\%          & 96.2\% \\
    0.33 & 0.33 & 0.33 & \textbf{98.2\%} & 99.0\%          & 95.9\%          & 97.2\%          & 98.5\%          & \textbf{96.6\%} \\
\bottomrule
\end{tabular}
\label{tab2}
\end{adjustbox}
\end{table}

\begin{table}[ht]
\centering
\caption{The results of the CETC model under 7 classes of $\beta$, $\alpha$, and $\gamma$ on the COVIDx CXR-3 dataset. The best values are highlighted in bold so you can easily identify them.}
\begin{adjustbox}{width=.95\linewidth,center}
\begin{tabular}[t]{ccccccccc}
\toprule
    $\alpha$    & $\beta$     & $\gamma$    & ACC             & NPV             & PPV             & SEN             & SPE             & FOS             \\
\midrule
    0.8         & 0.1         & 0.1         & 93.2\%          & 88.8\%          & 98.9\%          & 87.5\%          & 99.0\%          & 92.8\%          \\
    0.6         & 0.2         & 0.2         & 87.2\%          & 80.2\%          & 98.7\%          & 75.5\%          & 99.0\%          & 85.6\%          \\
    0.1         & 0.8         & 0.1         & 92.7\%          & 88.3\%          & 98.3\%          & 87.0\%          & 98.5\%          & 92.3\%          \\
    0.2         & 0.6         & 0.2         & 90.0\%          & 83.9\%          & 98.8\%          & 81.0\%          & 99.0\%          & 89.0\%          \\
    0.1         & 0.1         & 0.8         & 89.5\%          & 82.9\%          & 99.4\%          & 79.5\%          & \textbf{99.5\%} & 88.3\%          \\
    0.2         & 0.2         & 0.6         & 89.0\%          & 82.5\%          & 98.7\%          & 79.0\%          & 99.0\%          & 87.8\%          \\
    0.33 & 0.33 & 0.33 & \textbf{95.0\%} & \textbf{91.3\%} & \textbf{99.5\%} & \textbf{90.5\%} & \textbf{99.5\%} & \textbf{94.8\%} \\
\bottomrule
\end{tabular}
\label{tab3}
\end{adjustbox}
\end{table}

We conducted extensive experiments using CETC on the COVIDx CXR-3 dataset and the COVID-19 radiography dataset, exploring seven different combinations of $\beta$, $\alpha$, and $\gamma$ coefficients. Table \ref{tab2} shows the experimental results for the different datasets and Table \ref{tab3} shows the experimental outcomes for the different datasets. Our analysis of the COVID-19 radiography dataset revealed that the results of $\beta$, $\alpha$, and $\gamma$ had minimal effect on the model performance, as the evaluation metrics varied between 93.4\% and 99.2\%. Notably, when all three coefficients were set to 0.33, the model achieved the highest accuracy (ACC) of 98.2\% and the highest F-1 score (FOS) of 96.6\%. 
Conversely, when $\beta$, $\alpha$, and $\gamma$ were assigned values of 0.2, 0.6, and 0.2, respectively, the model attained the highest positive predictive value (PPV) of 97.5\% and the highest specificity (SPE) of 99.1\%. Additionally, the highest negative predictive value (NPV) of 99.2\% and sensitivity (SEN) of 97.8\% were illustrated when $\beta$, $\alpha$, and $\gamma$ were set to 0.6, 0.2, and 0.2. In Section \hyperref[4.2]{4.2}, we compare the model achievement and select the coefficient group of 0.33, 0.33, and 0.33 due to its highest ACC. On the other hand, in the case of the COVIDx CXR-3 dataset, we observed a significant impact of $\alpha$, $\beta$, and $\gamma$ on the model performance, resulting in evaluation metrics ranging from 75.5\% to 99.5\%. When all three coefficients were set to 0.33, the model achieved the highest values for NPV, ACC, PPV, SPE, SEN, and F-1 score (FOS), reaching 91.3\%, 95.0\%, 99.5\%, 99.5\%, 90.5\%, and 94.8\%, respectively.
However, when $\beta$, $\alpha$, and $\gamma$ were set to $0.1$, $0.1$, and $0.8$, the model demonstrated consistent specificity (SPE), whereas other metrics showed noticeable under performance. This notable discrepancy in performance across various coefficient groups can be traced back to differences in the distribution of feature scales within the dataset. In situations where the feature scale distribution is broad, features from all scales exert a considerable influence on the ultimate result. Therefore, optimal performance is attained when there is a proportional balance among $\alpha$, $\beta$, and $\gamma$. Hence, for the purpose of comparing model performance in Section \hyperref[4.2]{4.2}, we selected the coefficient group of 0.33, 0.33, and 0.33 due to its superior performance.

\subsection{Comparison with SOTA models}

The model performance of our CETC is compared to several SOTA models in this section. According to the results, we have shown that our CETC exceeds the SOTA methods by a wide margin for all six evaluation metrics on a range of levels. There is no doubt that our CETC exhibited slightly higher performance across all six metrics of the COVID-19 radiography dataset. It achieves an ACC of up to 98.2\% and the highest values for positive predictive value, sensitivity, negative predictive value, and F-1 score, and specificity , at 95.9\%, 99.0\%, 97.2\%, 96.6\%, and 98.5\%, respectively. The CETC report demonstrates significant superiority across various evaluation metrics, particularly in the areas of ACC, NPV, SEN, and FOS, for the COVIDx CXR-3 dataset. Notably, CETC exhibits a remarkable SEN leadership of around 30\%. The outstanding performance of CETC confirms the efficacy of the suggested model, as well as the importance of capturing both multi-local features as well as global features in the proposed architecture. The following tables provide detailed experimental results on the comparison of CETC to other SOTA models, as well as references to Table~\ref{tab4} and Table~\ref{tab5}.

\begin{table}[ht]
\centering
\caption{Performance comparing between the CETC and 7 SOTA models on the COVID-19 radiography dataset. $\beta$, $\alpha$, and $\gamma$ equals 0.33. The best values are shown in bold face.}
\begin{adjustbox}{width=.95\linewidth,center}
\begin{tabular}[t]{ccccccc}
\toprule
    Model               & ACC             & NPV             & PPV             & SEN             & SPE             & FOS    \\
\midrule
    VGGNet~\cite{simonyan2015very}         & 95.3\%          & 96.6\%          & 91.6\%          & 90.3\%          & 97.1\%          & 91.0\% \\
    ResNet~\cite{he2016deep}               & 92.9\%          & 93.2\%          & 92.0\%          & 79.8\%          & 97.5\%          & 85.5\% \\
    MobileNet~\cite{medvit}   & 97.0\%          & 98.3\%          & 93.2\%          & 95.3\%          & 97.5\%          & 94.3\% \\
    ConvNeXt~\cite{liu2022convnet}         & 95.7\%          & 95.8\%          & 95.2\%          & 87.8\%          & 98.4\%          & 91.4\% \\
    SwT~\cite{liu2021swin}                 & 96.2\%          & 96.4\%          & 95.6\%          & 89.8\%          & \textbf{98.5\%} & 92.6\% \\
    ViT~\cite{dosovitskiy2020image}        & 95.1\%          & 95.9\%          & 92.5\%          & 88.4\%          & 97.4\%          & 90.4\% \\
    MaxViT~\cite{tu2022maxvit}            & 91.5\%          & 91.5\%          & 91.2\%          & 74.6\%          & 97.4\%          & 82.1\% \\
    \textbf{CETC(Ours)} & \textbf{98.2\%} & \textbf{99.0\%} & \textbf{95.9\%} & \textbf{97.2\%} & \textbf{98.5\%} & \textbf{96.6\%} \\
\bottomrule
\end{tabular}
\label{tab4}
\end{adjustbox}
\end{table}

\begin{table}[ht]
\centering
\caption{Performance comparing between the CETC and 7 SOTA models on the COVIDx CXR-3 dataset. $\beta$, $\alpha$, and $\gamma$ equals "$0.33$". The best values are shown in bold face.}
\begin{adjustbox}{width=.95\linewidth,center}
\begin{tabular}[t]{ccccccc}
\toprule
    Model               & ACC             & NPV             & PPV             & SEN             & SPE             & FOS    \\
\midrule
    VGGNet~\cite{simonyan2015very}               & 77.5\%          & 69.9\%          & 94.4\%          & 58.5\%          & 96.5\%          & 72.2\% \\
    ResNet~\cite{he2016deep}             & 76.0\%          & 70.0\%          & 87.1\%          & 61.0\%          & 91.0\%          & 71.8\% \\
    MobileNet~\cite{medvit}           & 85.2\%          & 77.4\%          & 99.3\%          & 71.0\%          & \textbf{99.5\%} & 82.8\% \\
    ConvNeXt~\cite{liu2022convnet}           & 82.2\%          & 74.7\%          & 96.4\%          & 67.0\%          & 97.5\%          & 79.1\% \\
    SwT~\cite{liu2021swin}                 & 85.0\%          & 78.0\%          & 96.7\%          & 72.5\%          & 97.5\%          & 82.9\% \\
    ViT~\cite{dosovitskiy2020image}                  & 83.7\%          & 77.3\%          & 94.1\%          & 72.0\%          & 95.5\%          & 81.6\% \\
    MaxViT~\cite{tu2022maxvit}              & 75.0\%          & 69.1\%          & 86.2\%          & 59.5\%          & 90.5\%          & 70.4\% \\
    \textbf{CETC(Ours)} & \textbf{95.0\%} & \textbf{91.3\%} & \textbf{99.5\%} & \textbf{90.5\%} & \textbf{99.5\%} & \textbf{94.8\%} \\
\bottomrule
\end{tabular}
\label{tab5}
\end{adjustbox}
\end{table}

\section{Conclusion}

In this paper, we present the Controllable Ensemble Transformer and CNN network model for analysing medical images in a controlled way. Using the CETC model, you can obtain both high and low level features in medical images by combining the strengths of convolutional neural networks and transformers. It consists of three blocks, namely a CEB, a TDB, and a TCB. The CETC model conducted better in terms of various evaluation metrics than existing SOTA models based on experimental results conducted on publicly available COVID-19 datasets. This highlights the potential of the CETC model for accurate and efficient analysis of medical images. In our future endeavors, we aim to broaden the scope of the proposed method beyond the classification task, while simultaneously enhancing its capacity for feature extraction.

\bibliographystyle{IEEEtran}

\bibliography{article}

\end{document}